\documentclass[aps,twocolumn,superscriptaddress,prl,10pt]{revtex4-1}

\usepackage{bm}

\usepackage{graphicx}

\usepackage{subfigure}

\def\be{\begin{equation}}
\def\ee{\end{equation}}
\def\e#1{\label{#1}\end{equation}}
\def\bea{\begin{eqnarray}}
\def\eea{\end{eqnarray}}
\def\ea#1{\label{#1}\end{eqnarray}}

\def\bem#1{\begin{mathletters}\label{#1}}
\def\eml{\end{mathletters}}

\def\4#1{{\boldsymbol{#1}}}
\def\8#1{{\widetilde{#1}}}
\def\bse{\begin{subequations}}
\def\ese{\end{subequations}}

\begin{document}


\title{Spectroscopy of composite solid-state spin environments for \\improved metrology with spin ensembles}

\author{N. Bar-Gill}
\affiliation{Harvard-Smithsonian Center for Astrophysics, Cambridge, MA 02138, USA}
\affiliation{Department of Physics, Harvard University, Cambridge MA, USA}
\author{L. M. Pham}
\affiliation{School of Engineering and Applied Sciences, Harvard University, Cambridge, MA 02138, USA}
\author{C. Belthangady}
\author{D. Le Sage}
\affiliation{Harvard-Smithsonian Center for Astrophysics, Cambridge, MA 02138, USA}
\author{P. Cappellaro}
\affiliation{Department of Nuclear Science and Engineering, MIT, Cambridge MA, USA}
\author{J. R. Maze}
\affiliation{Department of Physics, Pontificia Universidad Catolica de Chile, Santiago 7820436, Chile}
\author{M. D. Lukin}
\author{A. Yacoby}
\affiliation{Department of Physics, Harvard University, Cambridge MA, USA}
\author{R. Walsworth}
\affiliation{Harvard-Smithsonian Center for Astrophysics, Cambridge, MA 02138, USA}
\affiliation{Department of Physics, Harvard University, Cambridge MA, USA}

\begin{abstract}
For precision coherent measurements with ensembles of quantum spins the relevant Figure-of-Merit (FOM) is the product of polarized spin density and coherence lifetime, which is generally limited by the dynamics of the spin environment.  Here, we apply a coherent spectroscopic technique to characterize the dynamics of the composite solid-state spin environment of Nitrogen-Vacancy (NV) centers in room temperature diamond. For samples of very different NV densities and impurity spin concentrations, we show that NV FOM values can be almost an order of magnitude larger than previously achieved in other room-temperature solid-state spin systems, and within an order of magnitude of the state-of-the-art atomic system. We also identify a new mechanism for suppression of electronic spin bath dynamics in the presence of a nuclear spin bath of sufficient concentration.  This suppression could inform efforts to further increase the FOM for solid-state spin ensemble metrology and collective quantum information processing.
\end{abstract}
%

\maketitle

Understanding and controlling the coherence of quantum spins in solid-state systems is crucial for precision metrology \cite{taylor2008,jero2008,gopi2008}, quantum information science \cite{jiang2009,neumann2010,mcCamey2010} and basic research on quantum many-body dynamics \cite{chuang,zurekRMP}.  Examples of such systems include Nitrogen-Vacancy (NV) color centers in diamond \cite{wrachtrup2006,childress2006,dutt2007}, phosphorous donors in silicon \cite{gere,tyryshkin} and quantum dots \cite{hansonRMP}. A Figure-of-Merit (FOM) for precision coherent measurements with quantum spins is the product of the polarized spin density ($n_{NV}$) and the coherence lifetime ($T_2$), with the phase-shift sensitivity $\delta \phi$ scaling as \cite{taylor2008,walsworth_imager,romalis}
%
\begin{equation}
\delta \phi \propto \frac{1}{\sqrt{n_{NV} T_2}} \equiv \frac{1}{\sqrt{FOM}}.
\end{equation}
Increasing this FOM requires an understanding of the sources of decoherence in the system, and their interplay with spin density.
For example, in the solid state $T_2$ is typically limited by interaction with an environment (i.e., bath) of paramagnetic spin impurities; whereas $n_{NV}$ is limited by fabrication issues, the associated creation of additional spin impurities in the environment, and the ability to polarize (state prepare) the quantum spins.

The 
paradigm of a central spin coupled to a spin environment has been studied intensively for many years (see e.g., \cite{dobrovitski,das_sarma2011}); and quantum control methods have been developed to extend the spin coherence lifetime by reducing the effective interaction with the environment. In particular, dynamical decoupling techniques pioneered in the field of NMR \cite{hahn,cp,cpmg,slichter} have recently been applied successfully to extend the effective 
$T_2$ of single NV-diamond electronic spins by more than an order of magnitude \cite{hanson2010,cory,hanson_bath_control}. 

\begin{figure}[htb]
\begin{center}
\subfigure[]{
\includegraphics[width=0.4 \linewidth,height=0.36 \linewidth]{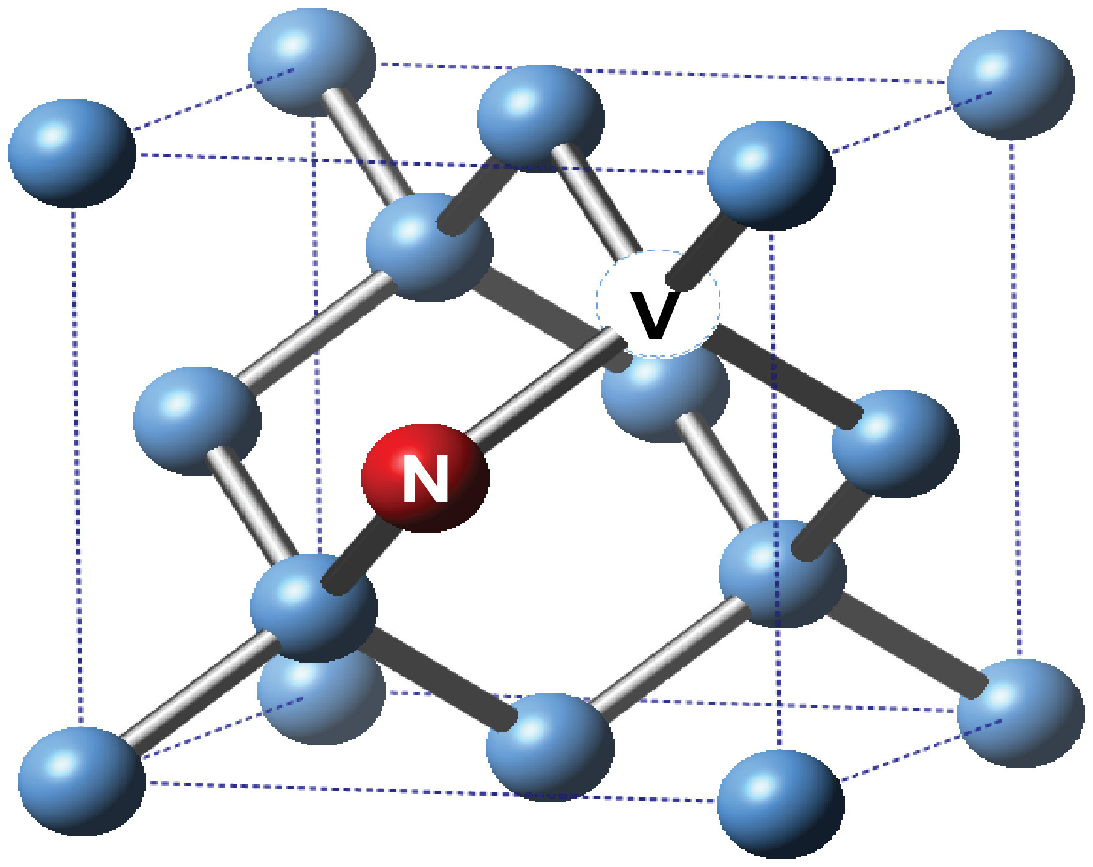}}
\subfigure[]{
\includegraphics[width=0.45 \linewidth,height=0.36 \linewidth]{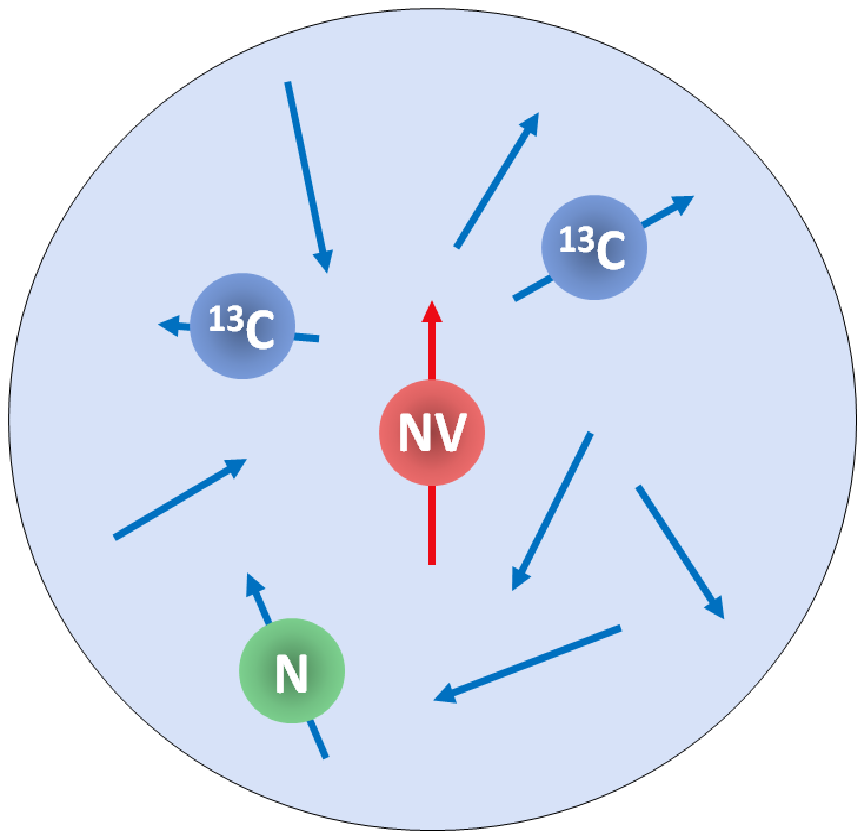}}
\subfigure[]{
\includegraphics[width=0.48 \linewidth,height=0.36 \linewidth]{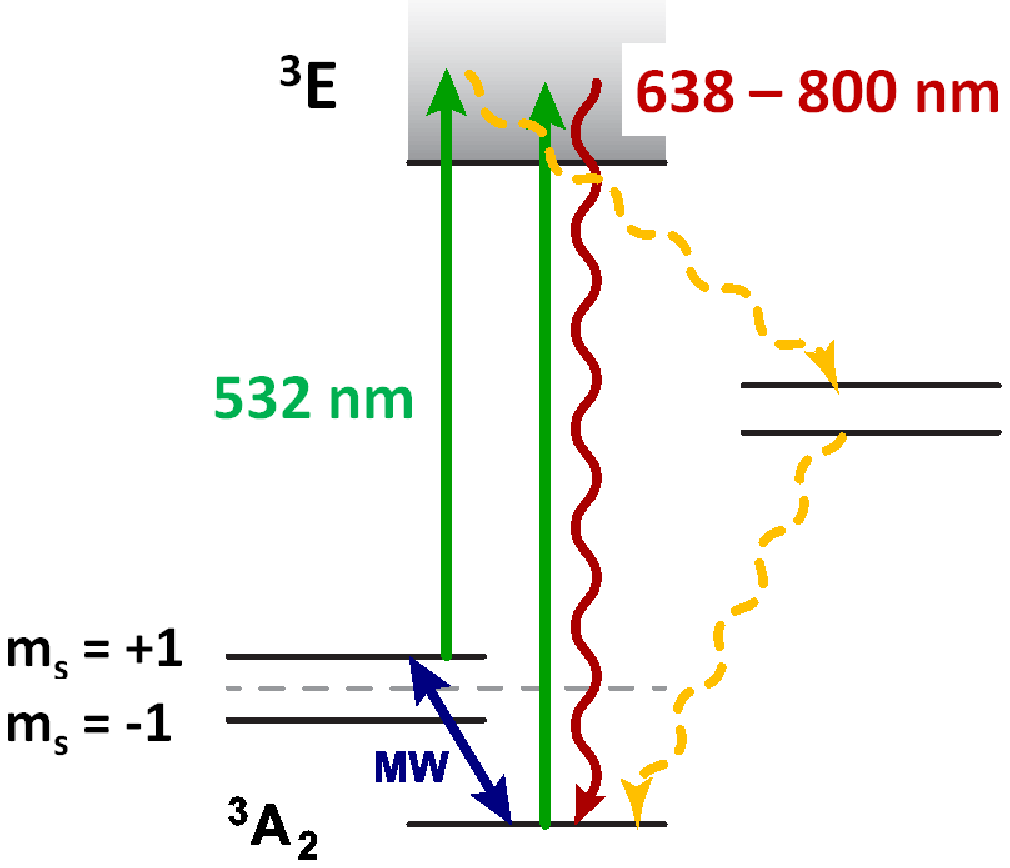}}
\subfigure[]{
\includegraphics[width=0.48 \linewidth,height=0.36 \linewidth]{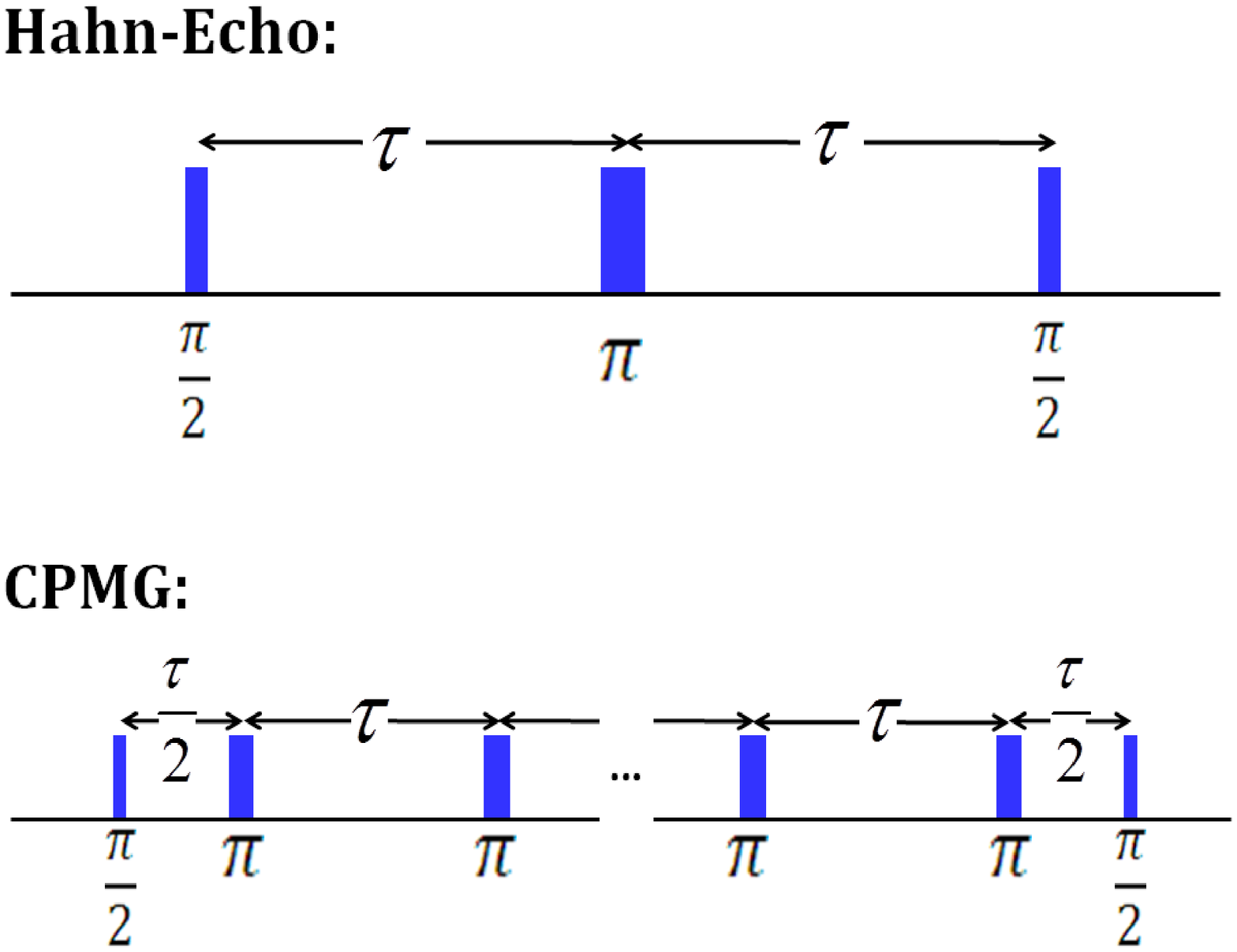}}
\protect\caption{Quantum spin in a solid-state environment (NV-center
in diamond) and the applied spin-control pulse-sequences. (a) Lattice
structure of diamond with a Nitrogen-Vacancy (NV) color center. (b) Magnetic
environment of the NV center electronic spin: $^{13}$C nuclear spin impurities
and N electronic spin impurities. (c) Energy-level schematic
of the NV center. (d) Hahn-echo and multi-pulse (CPMG)
spin-control sequences.} \label{fig:intro}
	\end{center}
\end{figure}

Here we study experimentally the FOM of ensembles of NV color centers in diamond (Fig. \ref{fig:intro}(a,b)) as a system for precision spin metrology such as magnetometry. To characterize the dynamics of the composite solid-state spin bath that limits the NV electronic spin coherence time, consisting of both electronic spin (N) and nuclear spin ($^{13}$C) impurities, we apply the technique of spectral decomposition \cite{almog,oliver}.
We describe this technique and its experimental application to three different diamond samples with a wide range of NV densities and impurity spin concentrations (see Table 1).  With the use of dynamical decoupling we realize very large values for the spin coherence $\mathrm{FOM} \sim 2 \times 10^{14} \mathrm{[ms/cm^3]}$ for two of these diamond samples, which is almost an order of magnitude larger than in other room-temperature solid-state systems \cite{liu_malonic}; three orders of magnitude greater than has been realized in NV-diamond systems for which $T_2$ has been measured \cite{walsworth_imager,wrachtrup_ultralong_2009}; and comparable to high-NV-density samples for which only the inhomogeneous coherence lifetime ($T_2^*$) was measured \cite{budkerNVT2}. In addition, the large NV FOM values realized here are only an order of magnitude smaller than the state-of-the-art alkali-vapor SERF (Spin Exchange Relaxation Free) magnetometer \cite{romalis}, which suffers from additional practical limitations compared to room-temperature solid-state devices.
We also find unexpectedly long correlation times for the electronic spin baths in two diamond samples with natural abundance of $^{13}$C nuclear spin impurities. We identify a new mechanism involving an interplay between the electronic and nuclear spin baths that can explain the observed suppression of electronic spin bath dynamics. In future work this effect could be used to engineer diamond samples for even larger NV spin FOM exceeding the current state-of-the-art, e.g., for precision magnetometry and collective quantum information applications.

The coherence of a two-level quantum system can be related to the magnitude of the off-diagonal elements of the system's density matrix. Specifically, we deal here with an ensemble of NV electronic spins in a finite external magnetic field, which can be treated as an effective two-level ensemble spin system with quantization (z) axis aligned with the nitrogen-vacancy axis.  When the NV spins are placed into a coherent superposition of spin eigenstates, e.g., aligned with 
the x-axis of the Bloch sphere, the measurable spin coherence is given by $C(t) =Tr [\rho(t) S_x ]$.
Due to coupling of the NV spins to their magnetic environment, 
the coherence is lost over time, with the general form
\begin{equation}
C(t) = e^{-\chi (t)},
\end{equation}
where the functional $\chi(t)$ describes the time dependence of the decoherence process. In the presence of a modulation acting on the NV spins (e.g., a resonant RF pulse sequence), described by a filter function in the frequency domain $F_t (\omega)$ (see below), it can be shown that the decoherence functional is given by \cite{klauder_anderson,universal_decay}
\begin{equation}
\chi(t) = \frac{1}{\pi} \int_0^{\infty} d \omega S(\omega) \frac{F_t (\omega)}{\omega^2}, \label{chi}
\end{equation}
where $S(\omega)$ is the spectral function describing the coupling of the system to the environment. Eq. (\ref{chi}) holds in the approximation of weak coupling of the NV spins to the environment, which is appropriate for the (dominantly) electronic spin baths of the diamond samples discussed here \cite{das_sarma2011} (see below).

The spectral decomposition technique allows us to determine $S(\omega)$ from straightforward decoherence measurements of an ensemble of NV spins, as outlined here. If we describe the magnitude of the coupling between the NV spins and the spin bath by a fluctuating energy term $\epsilon(t)$, then $S(\omega)$ is the Fourier transform of the second-order correlator of $\epsilon(t)$:
\begin{equation}
S(\omega) = \frac{1}{\hbar^2} \int_0^{\infty} d t e^{i \omega t} \langle \epsilon(0) \epsilon(t) \rangle.
\end{equation}

It can be seen from Eq. (\ref{chi}) that if an appropriate modulation is applied to the NV spins such that $F_t(\omega)/(\omega^2 t) = \delta (\omega- \omega_0)$, i.e., if a Dirac $\delta$-function (or close approximation) is localized at a desired frequency $\omega_0$, then $\chi(t) = t S(\omega_0)/\pi$. Therefore, by measuring the time-dependence of the spin ensemble's coherence $C(t)$ when subjected to such a ``spectral $\delta$-function'' modulation, we can extract the spin bath's spectral component at frequency $\omega_0$:
\begin{equation}
S(\omega_0) = -\pi \ln (C(t))/t.
\end{equation}
This procedure can then be repeated for different values of $\omega_0$ to provide complete spectral decomposition of the spin environment.

A close approximation to the ideal spectral filter function $F_t(\omega)$ described above can be provided by a variation on the well-known CPMG pulse sequence for dynamical decoupling of a spin system from its environment \cite{cpmg} (Fig. \ref{fig:intro}(d)). The CPMG pulse sequence is an extension of the Hahn-echo sequence \cite{hahn} (Fig. \ref{fig:intro}(d)), with $n$ equally-spaced $\pi$-pulses applied to the system after initially rotating it into the x-axis with a $\pi/2$-pulse. 
The resulting filter function for this sequence $F_{CPMG}(\omega)$ covers a narrow frequency region (given by $\pi/t$, where $t$ is total length of the sequence) which is centered at $\omega_0 = \pi n /t$, and is given by \cite{dassarma_dd}:
\begin{equation}
F^{CPMG}_n (\omega t) = 8 \sin^2 \left( \frac{\omega t}{2} \right) \frac{\sin^4 \left(\frac{\omega t}{4 n} \right)}{\cos^2 \left( \frac{\omega t}{2 n} \right)}.
\end{equation}
We note that the narrow-band feature of the CPMG filter essentially defines the effectiveness of this sequence for dynamical decoupling.
We apply a deconvolution procedure to correct for deviations of this filter function from the ideal Dirac $\delta$-function (see Supplement).

The composite solid-state spin environment of interest here, which is dominated by a bath of fluctuating N electronic spin impurities, causes decoherence of the probed NV spins through magnetic dipolar interactions. In the regime of low external magnetic fields and room temperature (relevant to the present experiments), the bath spins are randomly oriented, and their flip-flops (spin state exchanges) can be considered as random uncorrelated events \cite{das_sarma2011}. Therefore, the resulting spectrum of the bath's coupling to the NV spins can be reasonably assumed to be Lorentzian \cite{kittel,klauder_anderson}:
\begin{equation}
S (\omega) = \frac{\Delta^2 \tau_c}{\pi} \frac{1}{1 + (\omega \tau_c)^2}.
\end{equation}
This spin bath spectrum is characterized by two parameters: $\Delta$ is the average coupling strength of the bath to the probed NV spins; and $\tau_c$ is the correlation time of the bath spins with each other, which is related to their characteristic flip-flop time. In general, the coupling strength $\Delta$ is given by the average dipolar interaction energy between the bath spins and the NV spins, and the correlation time $\tau_c$ is given by the inverse of the dipolar interaction energy between neighboring bath-spins. Since such spin-spin interactions scale as $1/r^3$, where $r$ is the distance between spins, it is expected that the coupling strength scales as the bath spin density $n_{spin}$ (i.e., $\Delta \propto n_{spin}$), and the correlation time scales as the inverse of this density (i.e., $\tau_c \propto 1/n_{spin}$).

Note also that the multi-pulse CPMG sequence used in this spectral decomposition technique extends the NV spin coherence lifetime by suppressing the time-averaged coupling to the fluctuating spin environment. In general, the coherence lifetime $T_2$ increases with the number of pulses $n$  employed in the CPMG sequence. For a Lorentzian bath, in the limit of very short correlation times $\tau_c \ll T_2$, the sequence is inefficient and $T_2 \propto n^0$ (no improvement with number of pulses). In the opposite limit of very long correlation times $\tau_c \gg T_2$, the scaling is $T_2 \propto n^{2/3}$ \cite{slichter,taylor2008,hanson_magnetometry} (see also recent work on quantum dots \cite{marcus_dd_scaling}).

\begin{table}
\caption{Comparison of key characteristics and extracted ``average-fit'' Lorentzian spin bath parameters for the NV-diamond samples studied in this work. \label{table1}}
\begin{ruledtabular}
\begin{tabular}{l || l | l| l}
 & {\bf $^{12}$C} & {\bf Apollo} & {\bf HPHT} \\
\hline
Meas. technique & ensemble & ensemble & confocal \\
\hline
N concentration & $\sim 1$ ppm & $\sim 100$ ppm & $\sim 50$ ppm \\
\hline
NV density & $\sim 10^{14} \mathrm{[cm^{-3}]}$ & $\sim 10^{16} \mathrm{[cm^{-3}]}$ & $\sim 10^{12} \mathrm{[cm^{-3}]}$ \\
\hline
$^{13}$C concentration & $0.01 \%$ & $1.1 \%$ & $1.1 \%$ \\
\hline
Echo (1-pulse) $T_2$ & $240(6)$ $\mu$s & $2(1)$ $\mu$s & $5(1)$ $\mu$s \\
\hline
$T_2$ scaling & $n^{0.43(6)}$ & $n^{0.65(5)}$ & $n^{0.7(1)}$ \\
\hline
Max. achieved $T_2$ & $2.2$ ms & $20 \mu$s & $35 \mu$s \\
\hline
\hline
$\Delta$ (expected) & $\approx 60$ kHz & $\approx 6$ MHz & $\approx 3$ MHz \\
\hline
$\Delta$ (measured) & $30(10)$ kHz & $7(3)$ MHz & $1(1)$ MHz \\
\hline
$\tau_c$ (expected) & $\approx 15$ $\mu$s & $\approx 0.17$ $\mu$s & $\approx 0.34$ $\mu$s \\
\hline
$\tau_c$ (measured) & $10(5)$ $\mu$s & $3(2)$ $\mu$s & $10(5)$ $\mu$s \\
\hline
\hline
FOM $\mathrm{[ms/cm^3]}$ & $2 \times 10^{14}$ & $2 \times 10^{14}$ & $10^{10}$ \\
\end{tabular}
\end{ruledtabular}
\end{table}


Experimentally, we manipulate the $|0 \rangle$-$|1 \rangle$ spin manifold of the NV triplet electronic ground-state using a static magnetic field and microwave pulses; and employ a $532$ nm laser to initialize and provide optical readout of the NV spin states (Fig. \ref{fig:intro}(c); for details see Supplement and \cite{childress2006}).

We applied the spectral decomposition technique described above and extracted the spin bath parameters $\Delta$ and $\tau_c$ as well as the NV spin coherence FOM for three diamond samples with differing NV densities and concentrations of electronic and nuclear spin impurities (see Table I): (1) An isotopically pure $^{12}$C diamond sample grown with chemical vapor deposition (CVD) techniques (Element 6), was studied using an NV ensemble microscope \cite{walsworth_imager}. This sample has a very low concentration of $^{13}$C nuclear spin impurities ($0.01 \%$), a moderate concentration of N electronic spin impurities ($\sim 1$ ppm), and a moderate NV density ($\sim 10^{14} \mathrm{[cm^{-3}]}$). 
(2) A thin-layer CVD diamond sample (Apollo) with natural $^{13}$C concentration ($1.1 \%$), high N concentration ($\sim 100$ ppm), and large NV density ($\sim 10^{16} \mathrm{[cm^{-3}]}$) was also studied using the NV ensemble microscope. 
(3) A type 1b high-pressure high-temperature (HPHT) diamond sample (Element 6) with natural $^{13}$C concentration, high N concentration ($\sim 50$ ppm), and low NV density ($ \sim 10^{12} \mathrm{[cm^{-3}]}$) was studied using a confocal apparatus able to measure single NV centers. 

In Sample 1 ($^{12}$C) the overall spin environment for the probed NV spins is expected to be simply described by a bath of N electronic spin impurities. We studied this sample because of its combination of insignificant nuclear spin impurities and moderate electronic spin impurity concentration, suggesting a favorable (large) FOM value.
We selected Sample 2 (Apollo) for characterization in order to compare the NV spin coherence FOM at high N concentration and NV density with the FOM at moderate N concentration and NV density in the $^{12}$C sample.
Characterization of Sample 3 (HPHT) was important for confirming that similar electronic spin bath dynamics are determined by the spectral decomposition technique for single NV spins and NV ensembles in the high N concentration limit (even though we expect a small FOM value due to the low NV density of the HPHT sample).

Both the HPHT and Apollo samples have a natural $1.1\%$ abundance of $^{13}$C nuclear spin impurities: two orders of magnitude higher density than in the $^{12}$C diamond sample. Nonetheless, for the high N concentration in the HPHT and Apollo samples, the simple physical picture outlined above, of NV electronic spins coupled to independent N electronic and $^{13}$C nuclear spin baths, implies that the spin environment and NV spin decoherence should still be dominated by the bath of N electron spins.

\begin{figure}[htb]
\begin{center}
\subfigure[]{
\includegraphics[width=0.9 \linewidth,height=0.43 \linewidth]{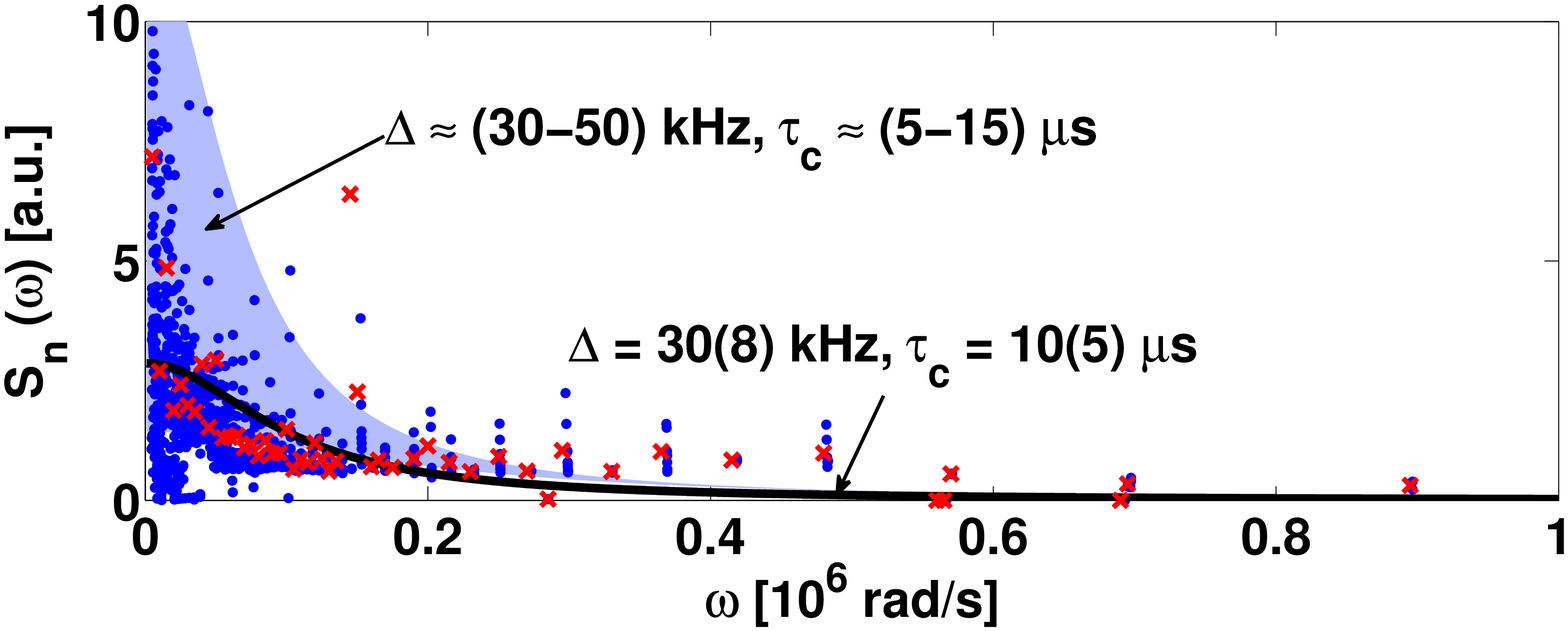}}
\subfigure[]{
\includegraphics[width=0.9 \linewidth,height=0.43 \linewidth]{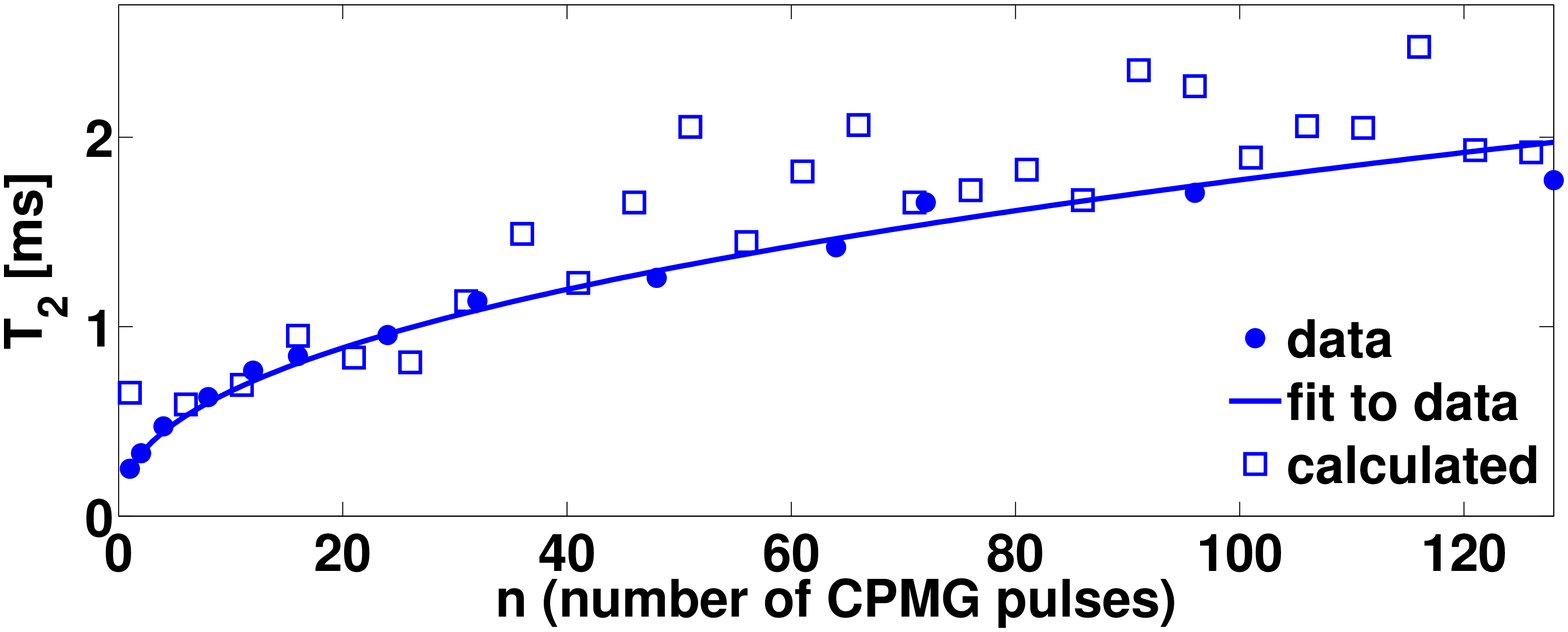}}
\protect\caption{Application of the spectral decomposition technique to an isotopically pure $^{12}$C NV-diamond sample. (a) Derived values for the spin bath spectral functions $S_n (\omega)$ for all CPMG pulse sequences (blue dots) and average values at each frequency (red crosses); best fit Lorentzian for the mean spectral function $\langle S_n(\omega) \rangle_n$ (solid black line); and range of best-fit Lorentzians for the individual spectral functions $S_n(\omega)$ for each CPMG pulse sequence (grey band). (b) Scaling of $T_2$ with the number $n$ of CPMG pulses: derived from NV spin coherence decay data $C_n(t)$ (dots); fit of dots to a power law $T_2 = 245 \mu s \times n^{0.43(6)}$ (solid line); and synthesized from the average-fit Lorentzian spin bath spectrum $\Delta=30$ kHz, $\tau_c=10$ $\mu$s (open squares).
} \label{fig:C12}
\end{center}
\end{figure}


The results for the $^{12}$C sample are summarized in Fig. \ref{fig:C12}.
As displayed in Fig. \ref{fig:C12}(a), we find that the best-fit Lorentzian spin bath spectrum (fitted to the average of all data points, see Supplement) has coupling strength of $\Delta = 30 \pm 10$ kHz and correlation time $\tau_c = 10 \pm 5$ $\mu$s, which agrees well with the range of values we find for the Lorentzian spin bath spectra $S_n (\omega)$ fit to each pulse sequence separately (see Supplement), $\Delta \simeq 30 - 50$ kHz, $\tau_c \simeq 5 - 15$ $\mu$s.
These values are also in reasonable agreement with the expected ``N dominated bath'' values for $\Delta$ and $\tau_c$, given this $^{12}$C NV-diamond sample's estimated concentrations of $^{13}$C and N spins (see Table I).

In Fig. \ref{fig:C12}(b) we plot the NV spin coherence lifetime $T_2$, determined from the measured coherence decay $C_n(t)$ for an $n$ pulse CPMG control sequence, as a function of the number of CPMG pulses out to $n=128$ (dots); and a power-law fit to this relationship, from which the scaling $T_2 \propto n^{0.43(6)}$ is found (solid line).
This power law exponent $<2/3$ is consistent with the spin bath $\tau_c$ being about an order of magnitude smaller than the Hahn-echo $T_2$ for the $^{12}$C diamond sample (see Table I).
The maximal coherence time achieved (using a $512$-pulse CPMG sequence) was $2.2$ ms (see Table I). 
To check the consistency of our data analysis procedure and assumptions, we used the best-fit Lorentzian spin bath spectrum discussed above as an input to calculate the $T_2(n)$ scaling by synthesizing an NV spin coherence decay $C_n(t)$ for each CPMG pulse sequence; and then fit to find $T_2$ for each synthesized $C_n(t)$ (empty squares in Fig. 2(b)). From this process we extract a scaling $T_2 \sim n^{0.4(1)}$, in good agreement with the scaling derived directly from the $C_n(t)$ data.



\begin{figure}[htb]
\begin{center}
\subfigure[]{
\includegraphics[width=0.9 \linewidth,height=0.43 \linewidth]{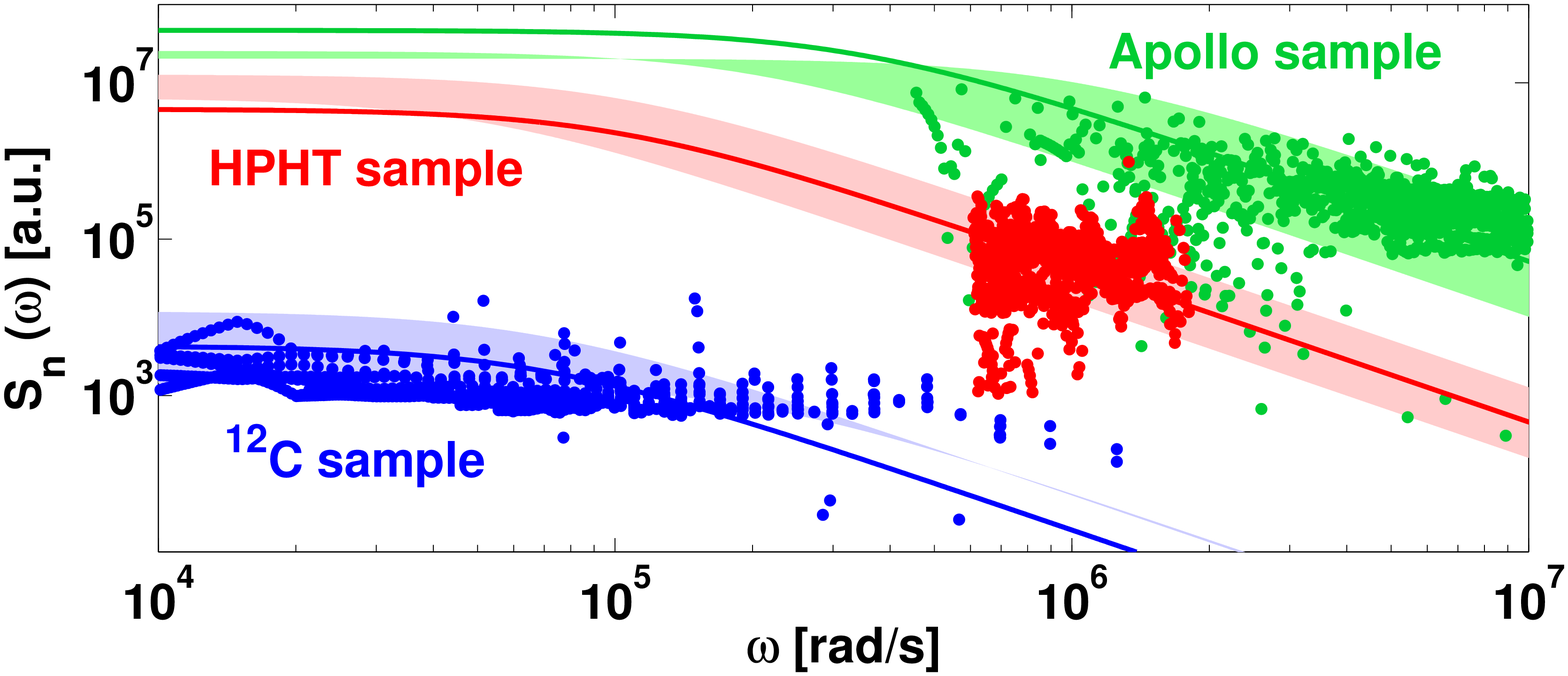}}
\subfigure[]{
\includegraphics[width=0.9 \linewidth,height=0.43 \linewidth]{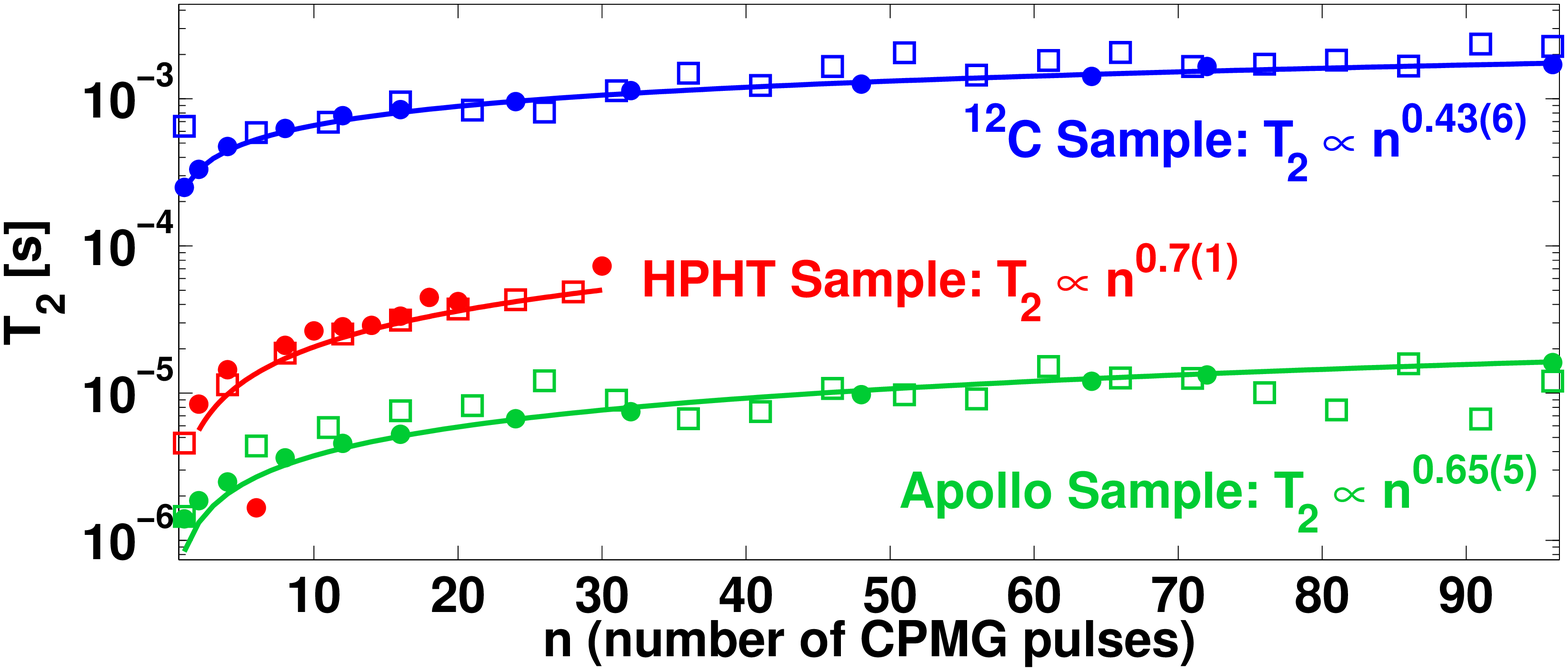}}
\protect\caption{Comparative application of the spectral decomposition technique to the $^{12}$C, HPHT and Apollo NV-diamond samples. (a) Spin bath spectral functions and associated Lorentzian fits using the average-fit method (solid lines) and the individual-fit method (color bands). (b) Scaling of $T_2$ with the number of CPMG pulses: dots - derived from NV spin coherence decay data $C_n(t)$; solid lines - fit of dots to a power law; open squares - synthesized from the average-fit Lorentzian spin bath spectrum.
} \label{fig:comparison}
\end{center}
\end{figure}

The results of the spectral decomposition procedure for all three samples are displayed in Fig. \ref{fig:comparison} and Table \ref{table1}. Fig. \ref{fig:comparison}(a) plots the derived spin bath spectral functions $S_n(\omega)$ and associated Lorentzian fits; 
and Fig. \ref{fig:comparison}(b) shows the measured scaling of $T_2$ with the number of CPMG pulses. From the full set of data for each sample we extract the coupling strength and correlation time of the best average-fit Lorentzian, which we verified to recreate correctly the decoherence curves and the $T_2$ scaling, using the synthesized data analysis procedure outlined above. 

Importantly, as listed in the bottom row of Table I, we found very large values for the spin coherence FOM $\sim 2 \times 10^{14} \mathrm{[ms / cm^3]}$ for the $^{12}$C and Apollo samples. These FOM values are 3 orders of magnitude larger than previously obtained in other room temperature NV diamond systems for which $T_2$ has been measured, with either single spins \cite{wrachtrup_ultralong_2009} or ensembles \cite{walsworth_imager}; and comparable to the FOM for a high-NV-density sample for which only $T_2^*$ was measured \cite{budkerNVT2}. The present results should inform improved precision spin metrology, e.g., magnetometry using NV diamond. The similar large FOM values for the $^{12}$C and Apollo samples are qualitatively consistent with the trade-off of increased NV density causing correspondingly decreased $T_2$ due to increased N impurity concentration, for a similar N-to-NV conversion efficiency $\sim 0.1 \%$ for these CVD samples. The HPHT sample has very low N-to-NV conversion efficiency $\sim 10^{-7}$, and hence has a much smaller FOM.

Also as seen in Table \ref{table1}, there is reasonable agreement between the measured and expected values for the coupling strength $\Delta$ in all three NV-diamond samples, with $\Delta$ scaling approximately linearly with the N concentration.  Similarly, the measured and expected values for the correlation time $\tau_c$ agree well for the $^{12}$C sample with low N concentration.  However, we find a striking discrepancy between the measured and expected values of $\tau_c$ for the two samples with $1.1\%$ $^{13}$C concentration and high N concentration (HPHT and Apollo): both these samples have measured spin bath correlation times that are more than an order of magnitude longer than given by the above simple electronic spin bath model, though the relative values of $\tau_c$ for the HPHT and Apollo samples scale inversely with N concentration, as expected.

The qualitative agreement between the results of the Apollo and HPHT samples indicates that the measured behavior of the electronic spin bath at high N concentration is not an artifact of the few NVs measured in the HPHT sample, and is not related to ensemble averaging in the Apollo sample.



The surprisingly long spin bath correlation times for the $^{13}$C diamond samples, revealed by the spectral decomposition technique, imply a suppression of the electronic (N) spin bath dynamics at higher nuclear spin ($^{13}$C) concentration.  We explain this suppression as a result of the random, relative detuning of electronic spin energy levels due to interactions between proximal electronic and nuclear spin impurities.
The $^{13}$C nuclear spins create a local Overhauser field for the N electronic spins, which detunes the energy levels of neighboring N spins from each other, depending on their (random) relative proximity to nearby $^{13}$C spins \cite{portis}. This detuning then suppresses the flip-flops of N spins due to energy conservation, effectively increasing the correlation time of the electronic spin bath in the presence of a finite concentration of nuclear spin impurities (compared to the correlation time in the $^{12}$C sample). 

The ensemble average effect of the composite electronic-nuclear spin bath interaction is to induce an inhomogeneous broadening $\Delta E$ of the N resonant electronic spin transition.
The electronic spin flip-flop rate $R$ is modified by this inhomogeneous broadening according to the expression \cite{sousa2003,witzel}
\begin{equation}
R \simeq \frac{\pi}{9} \frac{\Delta_{N}^2}{\Delta E},
\end{equation}
where $\Delta_{N}$ is the dipolar interaction between N electronic spins.  In this physical picture, $\Delta E$ is proportional to the concentration of $^{13}$C impurities and to the hyperfine interaction energy between the N electronic and $^{13}$C nuclear spins; whereas $\Delta_{N}$ is proportional to the N concentration. Taking into account the magnetic moments and concentrations of the N electronic and $^{13}$C nuclear spin impurities in our samples, and the large contact hyperfine interaction for N impurities in diamond \cite{cox,guven}, we estimate $\Delta E \sim 10$ MHz and $\Delta_{N} \sim 1$ MHz for the HPHT and Apollo samples.  This estimate implies approximately an order of magnitude suppression of $R$ compared to the bare flip-flop rate ($R_{bare}\sim \Delta_{N}$), which is consistent with the long spin bath correlation times for the $^{13}$C diamond samples, as revealed by the spectral decomposition technique (see Table \ref{table1}).


In summary, we applied a spectral decomposition technique to three NV-diamond samples with different composite-spin environments to characterize the bath dynamics and determine the NV spin coherence Figure-of-Merit (FOM).
Using this technique we realized for two samples a FOM $\sim 2 \times 10^{14} \mathrm{[ms/cm^3]}$, which is almost an order of magnitude larger than in other room-temperature solid-state spin system; and within one order of magnitude of the largest spin coherence FOM achieved in atomic systems -- an alkali vapor SERF magnetometer \cite{romalis} (the highest sensitivity magnetometer to date). The effective NV spin coherence time achieved with the $^{12}$C sample ($2.2$ ms for $512$ CPMG pulses) was nearly $T_1$ limited and comparable to the best coherence times achieved with single NV spins, suggesting possible applications in collective quantum information processing.
For samples with a finite concentration of $^{13}$C nuclear spin impurities, this technique revealed a significant suppression of the N electronic spin bath dynamics by interactions between the nuclear and electronic spin baths.
Further optimization of the NV spin coherence FOM may be possible by engineering this suppression, e.g., with $^{13}$C concentration higher than the natural value. 
This possibility is in contrast to the alkali vapor SERF FOM, which is fundamentally limited as atomic density is increased by spin destruction collisions between atoms \cite{romalis}, and therefore holds the promise of NV-diamond devices exceeding the current state-of-the-art.

Finally, the spectral decomposition technique presented here, based on well-known pulse sequences and a simple reconstruction algorithm, can be applied to other composite solid-state spin systems, such as quantum dots and phosphorous donors in silicon. Such measurements could provide a powerful approach for the study of many-body dynamics of complex spin environments, and enable identification of optimal parameter regimes for a wide range of applications in quantum information science and metrology.

We gratefully acknowledge fruitful discussions with Patrick Maletinsky and Shimon Kolkowitz; and assistance with samples by Daniel Twitchen and Matthew Markham (Element 6) and Patrick Doering and Robert Linares (Apollo). This work was supported by NIST, NSF and DARPA (QuEST and QuASAR programs).


\end{document}